\documentstyle[pre,aps,epsfig,twocolumn]{revtex}
\draft
\newcommand{\newc}{\newcommand}
\newc{\beq}{\begin{equation}}
\newc{\eeq}{\end{equation}}
\newc{\kt}{\rangle}
\newc{\br}{\langle}
\newc{\longra}{\longrightarrow}
\newc{\eps}{\epsilon}

\begin{document}

\title{Phase space localization of chaotic eigenstates: Violating
ergodicity}

\author{Arul Lakshminarayan$^{1,2}$, Nicholas R. Cerruti$^{1}$ and Steven
Tomsovic$^{1}$}
\address{$^{1}$ Dept. of Physics, Washington State University,
Pullman, WA 99164-2814 USA}
\address{$^{2}$ Physical Research Laboratory, Navrangpura, Ahmedabad 380
009, India.}

\date{\today}
\maketitle

\begin{abstract}
The correlation between level velocities and eigenfunction intensities
provides a new way of exploring phase space localization in quantized
non-integrable systems. It can also serve as a measure of deviations from
ergodicity due to quantum effects for typical observables.  This paper
relies on two well known paradigms of quantum chaos, the bakers map and
the standard map, to study correlations in simple, yet chaotic,
dynamical systems.  The behaviors are dominated by the presence of
several classical structures. These primarily include short periodic
orbits and their homoclinic excursions. The dependences of the
correlations deriving from perturbations allow for eigenfunction features
violating ergodicity to be selectively highlighted.  A semiclassical
theory based on periodic orbit sums leads to certain classical
correlations that are super-exponentially cut off beyond a logarithmic
time scale.  The theory is seen to be quite successful in reproducing
many of the quantum localization features.

\end{abstract}


\section{Introduction}

For a bounded, classically chaotic system, ergodicity is defined with
respect to the energy surface, the only available invariant space
of finite measure.  In an extension developed just over twenty years ago,
the consequences of ergodicity for the eigenstates of a corresponding
quantum system were conjectured to give rise to a locally, Gaussian
random behavior~\cite{Berry77,Voros79}.  Shortly thereafter, work ensued
on defining the concept of eigenstate ergodicity within a more rigorous
framework~\cite{stechel}.  Some of the paradoxes and peculiarities have
been recently explored as well~\cite{kaplan1}.  One expression of
eigenstate ergodicity is that a typical eigenstate would fluctuate over
the energy surface, but otherwise be featureless, in an appropriate
pseudo-phase-space representation such as the Wigner transform
representation~\cite{balazs}.  Any statistically significant deviation
from ergodicity in individual eigenstates is termed phase space
localization.

It came as a surprise when Heller discovered eigenstates were ``scarred''
by short, unstable periodic orbits~\cite{Heller84,LesHou91}.
A great deal of theoretical and numerical research
followed~\cite{Heller86,Bogom88,Berry89,EHP89,Sarac90,SSA98}, and
experiments also~\cite{From94,Muller95}.  In fact, scarring is just one of
the means by which phase space localization can exist in the eigenstates
of such systems.  Another means would be localizing effects due to
transport barriers such as cantori~\cite{mckay,ketzmerick} or broken 
separatrices~\cite{btu}.  Despite these studies and the semiclassical 
construction of an eigenstate~\cite{th}, the properties of individual 
eigenstates remain somewhat a mystery.

Individual eigenfunctions may not be physically very relevant in
many situations, especially those involving a high density of states.
In this case, groups of states contribute towards localization in
ways that may be understood with available semiclassical theories. One
simple and important quantity where this could arise is the
time average of an observable as this is a weighted sum of several
(in principle all) eigenfunctions.  In the Heisenberg picture, where the
operator is evolving in time, the expectation value of the observable
could be measured with any state.  Phase space localization
features would be especially evident if this state were chosen to be a
wave packet well localized in such spaces.

One of us has already studied the correlation between level velocities and
wavefunction intensities in connection with localization \cite{Steve96}.
This can be directly connected to the issues raised above, and our
treatment thus extends the previous work.  The present paper is a
companion to a study of similar problems in continuous Hamiltonian
systems (as opposed to Hamiltonian maps, i.e. discrete time systems) and
billiards~\cite{Nick}.  The methods used in these two papers complement
each other and the results in the present paper are detailed as the
systems studied are much simpler.

In a system with a non-degenerate spectrum the time average of an
observable $\hat A$ in state $|\alpha \kt$ is
\beq
\label{opavg}
\left< \hat A(t) \right>_t \, =\, \sum_{n} |\br \alpha|\psi_n \kt|^2
\br \psi_n|\hat A|\psi_n \kt, \eeq
where $|\psi_n \kt$ are the eigenstates of the Hamiltonian $\hat H$.  If
the system depends upon a parameter $\lambda$ which varies continuously,
an energy level ``velocity'', $\partial E_n/\partial \lambda$, for the
$n^{th}$ level can be defined (velocity is a bit of a misnomer for it is
actually just a slope - we are not evolving the system parameter in 
time).  The level~velocity-intensity correlation
is identical to the time average above if we identify $\partial \hat
H/\partial \lambda$ with $\hat A$, as from the so called Hellmann-Feynman
theorem (for instance~\cite{LangeRaab}):
\beq \br \psi_n |\frac{\partial \hat H}{\partial \lambda} |\psi_n
\kt\, =\, \frac{\partial E_n}{\partial \lambda}.  \eeq It must be
noted that while we call the operator average Eq. (\ref{opavg}) a
``correlation'' it is not the true correlation that is obtained by
dividing out the rms values of the wavefunction intensities and the
operator expectation value (as defined in \cite{Steve96}).  In other
words we are going to study the covariance rather than the
correlation. This is followed in this paper for two reasons;
first, dividing out these quantities does not retain the meaning of
the time average of an observable and second, the root mean square of the
wavefunction intensities which is essentially the inverse
participation ratio in phase space is itself a fairly complex quantity
reflecting on phase-space localization.

\section{ Correlation for the quantum bakers map}

\subsection{Generalities}

We first formulate in general the quantity to be studied and the
approach to its semiclassical evaluation. The classical dynamical
systems are discrete maps on the dimensionless unit two-torus whose
cyclical coordinates are denoted $(q,p)$. The quantum kinematics
is set in
a space of dimension $N$ \cite{schwinger,Berry80,BalVor89}
where this is related to the scaled Planck constant
as $N=1/h$, and the classical limit is the large $N$ limit. The quantum
dynamics is
specified by a unitary operator $U$ (quantum map)
that propagates  states by one discrete
time step. The quantum stationary states are the eigensolutions of this
operator. The $N$ eigenfunctions and eigenangles  are denoted by
$\{|\psi_i \kt\,, \, \phi_i ;\; i=0, \cdots, N-1\}$.
The eigenvalues lie on the unit circle and are members of the set
$\{\exp(-i \phi_i)\; i=0, \cdots, N-1\}$.

The central object of interest is:
\beq
\label{correl}
C_A(\alpha)\, =\, \sum_{i=0}^{N-1} |\br \alpha| \psi_i \kt |^2 \br \psi_i 
|\hat{A}| \psi_i \kt.
\eeq
The operator $\hat{A}$ is a  Hermitian operator and either describes an
observable or a perturbation allowing one to follow the levels'
motions continuously. The state $|\alpha \kt$ represents a  wave packet
on the two-torus that  is well localized in the $(q,p)$ coordinates
\cite{Sarac90,CTH92}. We will be interested in the quantum effects over
and above the classical limit and we will require that the operator is
traceless. Otherwise we will need to subtract the uncorrelated product of
the averages of the  eigenfunctions (unity) and the trace of the
operator. This immediately also implies that the correlation according to
Random Matrix Theory (RMT)~\cite{PandeyRMT} is zero as well. The ensemble
average of $C_A(\alpha)$ will wash out random oscillations that are a
characteristic of the Gaussian distributed  eigenfunctions of the random
matrices. Specific localization properties that we will discuss are then
not part of the RMT models of quantized chaotic  systems. In the
framework of level velocities we are considering the  situation where the
average level velocity is zero, {\em i.e.,} there is  no net drift of the
levels.

As we noted in the introduction the correlation $C_{A}(\alpha)$ is
simply the time average of the observable:
\beq
\label{opavgmap}
C_A(\alpha)\, =\, \left< \br \alpha |U^{-n} \hat{A} U^{n}| \alpha \kt \right>_n
\eeq
where the large angular brackets denotes the time $(n)$ average.
This requires that degeneracies do not exist and we assume that
this is the case as we are primarily interested in quantized
chaotic systems.
A physically less transparent identity that is nevertheless useful
in subsequent evaluations is:
\beq
\label{obscure}
C_A(\alpha)\, =\, \left<  \br \alpha |U^n|\alpha \kt  \, \mbox{Tr}
\left( \hat{A} U^{-n} \right) \, \right> _n
\eeq
This may be written more symmetrically as
\beq
C_A(\alpha)\, =\,   \left< \mbox{Tr}
\left( |\alpha \kt \br \alpha | U^{n} \right)  \, \mbox{Tr}
\left( \hat{A} U^{-n} \right) \, \right> _n.
\eeq
Thus, the correlation is a sort of time evolved average correlation
between the two operators $\hat{A}$ and $|\alpha \kt \br \alpha |$. The
semiclassical expressions for these are however different as
complications arise from the classical limit of $|\alpha \kt \br \alpha
|$ which would be varying over scales of $\hbar$ that govern the validity
of the stationary phase approximations.  However, we may anticipate,
based on the last form, that the semiclassical expression would be
roughly the correlations of the classical limits of these two
operators~\cite{ANS99}.

\subsection{Semiclassical Evaluation}

The bakers map is a very attractive system to study the quantities
discussed in the introduction. The classical dynamics is particularly
simple (it is sometimes referred to as the  ``harmonic oscillator of chaos'').
A simple quantization is due to Balazs and Voros \cite{BalVor89} (where a
discussion of the classical dynamics may also found). As a model of
quantum chaos it shows many generic features including the  one central
to this study namely scarring localization of
eigenfunctions~\cite{Sarac90}.  There are detailed semiclassical theories
that have been verified
substantially~\cite{CTH92,OT91,AlmSarac91,SaracVor}.  We neglect certain
anomalous features of the quantum bakers map~\cite{OT91,SaracVor} that
would eventually show up in the classical limit.  This is reasonable in
the range of scaled Planck constant values we have used in the following.

We use the second time averaged expression, Eq. (\ref{obscure}), for the
correlation.  We do not repeat here details of the quantization of the
bakers map or the semiclassical theories of this operator except note
that we use the anti-periodic boundary conditions as stipulated by
Saraceno \cite{Sarac90} in order to retain fully the classical
symmetries.

The semiclassical theory of the bakers map deals with the powers of
the propagator.  The trace of $U^n$, the time $n$ propagator, has been
written in the canonical form of a sum over classical hyperbolic
periodic orbits with the phases being actions and the amplitudes
relating to the linear stability of the orbits.  The complications with
Maslov phases is absent here \cite{OT91,AlmSarac91}.  Also, the
semiclassical expressions have been derived for matrix elements of the
time $n$ propagator in the wave packets basis \cite{CTH92}.  The
time domain dominates the study of the quantum maps, the Fourier
transform to the spectrum being done exactly.  Our approach to the
correlation is then naturally built in the time domain.  The situation
is different in the case of Hamiltonian time independent flows where
the energy domain is very useful.

We use the semiclassical expression for the propagator diagonal matrix
elements derived in \cite{CTH92}:
\begin{eqnarray}
\label{semprop}
\br \alpha |U^n|\alpha \kt \, &\sim& \, \sum_{\gamma} \frac{\exp(i 
S_{\gamma}/\hbar)}
{\sqrt{\cosh(\lambda n )}} \sum_j
\exp \left[ -\frac{\cosh(n \lambda)-1}
{2 \cosh(n \lambda) \hbar} \right. \nonumber \\
&& \left. \times ( \delta q^2 +\delta p^2 ) - \frac
{i \delta q \delta p}{\hbar} \tanh(n \lambda) \right].
\end{eqnarray}
Here $\gamma$ labels periodic orbits of period $n$ including repetitions.
The Lyapunov exponent is $\lambda$ which is  $\ln(2)$ for the usual bakers
map (corresponding to the $(1/2,1/2)$ partition and Bernoulli
process). Also $ \hbar= h/(2 \pi) = 1/( 2 \pi N)$,
$ \delta q = q_j -q_{\alpha}$ and a similar relation for $ p$.
The position of $j$-th periodic point on the periodic
orbit $\gamma$ is $(q_j,p_j)$.  The centroids of the wave packets, 
assumed circular Gaussians, are $(q_{\alpha},p_{\alpha})$.
The choice of type of wave packets is not crucial for the features we
seek. We note that the simplicity of this expression for the propagator
derives from the simplicity of the classical bakers map, especially the
fact that the stable and unstable manifolds are everywhere aligned with
the $(q,p)$ axes. That Eq. (\ref{semprop}) happens to be a
periodic orbit sum differs from the similar treatment for billiards
as found in \cite{SteveHell93} where such sums are treated as homoclinic
orbit sums.  Note however, that the local linearity of the bakers map
renders the two approaches (periodic orbit, homoclinic orbit) equivalent.

A generalization of the trace formula for the propagator is given below
that is easily derived by the usual procedure employed for the
propagator itself \cite{AlmSarac91}.  Such a formula was derived in
\cite{EFMW92} for the case of Hamiltonian flows in the energy domain. We
make  the simplifying assumption that the operator $\hat{A}$ is diagonal in the
position representation (we could treat the case of $\hat{A}$ being diagonal
in momentum alone as well). This avoids the problem of a Weyl-Wigner
association of operators to functions on the torus. The quantum operator
$\hat{A}$ under this simplifying assumption has an obvious classical limit
and associated function which is denoted by $A(q)$. The other
major assumption used in deriving the formula below is that it does
not vary on scales comparable to or smaller than $\hbar$.

Thus we derive:
\beq
\label{semtrace}
\mbox{Tr}\left( \hat{A} U^{-n} \right) \, \sim \, \sum_{\gamma} 
\frac{\exp(-i S_{\gamma}/\hbar)}
{2 \sinh( n \lambda/2)} \, \sum_{j} A(q_j).
\eeq
The index $j$ again labels points along the periodic orbit $\gamma$.
The sum over the periodic orbit is the analogue of the integral
of the Weyl transform over a primitive periodic orbit in the
Hamiltonian flow case \cite{EFMW92}.
The special case $\hat{A}=I$ the identity corresponds to the usual trace
formula \cite{OT91,AlmSarac91}. Note that we have written the sums above
as being over periodic orbits, while the trace formulas have
been often written as sums over {\em fixed points}.

The first step is to multiply the two semiclassical periodic orbit sums in
Eq. (\ref{semprop}) and in Eq. (\ref{semtrace}).  Since there is a
time average, $n$ is assumed large enough, but not too large (so
that these expansions retain some accuracy).  All hyperbolic
functions are approximated by their dominant exponential dependences.  The
diagonal approximation and the uniformity principle \cite{HannAlm} is
used as well.
\begin{eqnarray}
C_A(\alpha) \, &=&\, \left<   \sum_{\gamma} \sqrt{2} \exp(-n\lambda)
\sum_{T} \left( \sum_j F(q_j, p_j) \right) \right. \nonumber \\
&& \left. \times \left( \sum_j  A(T q_j, T p_j) \right) \right> _n
\end{eqnarray}
Here we have taken a more general dependence for $A$ (including
the possibility of momentum dependence).  $T$ represents elements of the
symmetry group of the system including time-reversal symmetry and
including, of course, unity. These symmetries imply in general, though
not as a rule, distinct (for $T \ne I$) orbits
with identical actions. One assumes that the overwhelming number of
action degeneracies are due to such symmetries.

The function $F$ is the approximated Gaussian:
\beq
\label{bakerF}
F(q_j, p_j) \, =\,
\exp \left[ -\frac{1}
{2 \hbar}( \delta q^2 +\delta p^2 ) - \frac
{i \delta q \delta p}{\hbar} \right]
\eeq
Using $\lambda=\ln(2)$ and the fact that there are approximately
$2^n/n$ orbits of period $n$, one finds
\beq
\label{fincor}
C_A(\alpha)\, =\, \sqrt{2} \sum_{T} \sum_{l= -M }^{M} \tilde C_T(l)
\eeq
where $\tilde C(l)$ is a classical $l$-step correlation:
\beq
\label{classcor}
\tilde C_T(l)=\frac{1}{n} \sum_{j=1}^{n} F(q_j, p_j) A(T q_{j+l}, T p_{j+l}).
\eeq
The time average is taken over a {\em typical} orbit. We abandon any specific
periodic orbit and appeal to ergodicity, taking $n$ and also $M$ as practically
infinite. This is with the assumption that such correlations will decay with
time $l$. In fact, below we calculate such correlations explicitly and
display the decay. Note that $\tilde C_T(l) \ne \tilde C_T(-l)$ in general.
Although these are classical correlations, in the sense that $q_j, p_j$
represent a classical orbit, $\hbar$ appears as
a parameter in them through $F$.
Further, using the ergodic principle we can replace time averages in
$\tilde C_T(l)$ by
appropriate phase space averages:
\beq
\tilde C_T(l) \, =\, \int dq \, \int  dp \, F(q,p) A(T f^l (q,p), T g^l(q,p))
\eeq
where we have used the fact that the  total phase space volume (area)
is unity, and $f^l(q,p)=q_l$,
$g^l(q,p)=p_l$ are the classical $l$-step integrated mappings.

\subsection{ Special case and verifications}

We first consider the case that $\hat{A} = A_0(\hat{T}_p 
+ \hat{T}_p ^{\dagger})/2$, where $\hat{T}_p$ is the unitary 
single-step momentum translation operator that is diagonal in 
the position representation and $A_0$ is a constant real 
number.  This implies that the associated function is 
$A(q)=A_0\cos( 2\pi q )$.  Below we consider $A_0=1$ as the strength of 
the perturbation.  The elements of $T$, apart from the identity ($I$), 
are time-reversal ($TR$) symmetry and parity ($P$). Time reversal 
in the bakers map is $(T(q) =p, T(p) =q)$ followed by backward iteration, 
while parity is the transformation $(T(q)=1-q, T(p)=1-p)$.

We begin with the evaluation of the forward correlation $(l \ge 0)$
corresponding to $T=I$.
\beq
\tilde C_I(l) \, =\, \int_{-\infty}^{\infty} dq \, dp \, F(q,p) 
\cos(2 \pi 2^l q).
\eeq
This follows from the equality:
\beq
f^l(q)= 2^l q (\mbox{mod} \, 1)
\eeq
for the bakers map.  The limits of the integrals can be extended to the
entire plane as long as the centroid of the weighting factor
$(q_\alpha,p_\alpha)$ is far enough away from the edges of the unit
phase space square that the Gaussian tails are small there.  The
integral is elementary, and using
$h=1/N$ one gets:
\beq
\label{forcor}
\tilde C_I(l \ge 0)\, =\, \frac{1}{\sqrt{2} N}
\exp( -2^{2l} \pi /(2N)) \cos(2 \pi 2^l q_\alpha).
\eeq
This explicit expression shows the super-exponential decrease with time
$l$ in the correlation coefficients.  It is interesting to note that the
logarithmic time scale which sets an important quantum-classical
correspondence scale of divergence for chaotic systems, here
$\tau=1/\lambda\ \ln(1/2\pi\hbar)=\ln(N)/\ln(2)$, enters the correlation
decay.  In fact, the correlations are significant to precisely half the
log-time.  We anticipate this feature to hold in general, including
autonomous Hamiltonian systems.

Since $g^{-l}(p)\, =\, 2^l p (\mbox{mod }\, 1)$, for $(l \ge 0)$,
the time-reversed
backward correlation $(l\le0)$ is
\beq
\tilde C_{TR}(l \le 0)\, =\,\frac{1}{\sqrt{2} N} \exp( -2^{-2l} \pi /(2N))
\cos(2 \pi 2^{-l} p_\alpha)
\eeq
which also decays super-exponentially and is responsible for the
$(q \leftrightarrow  p)$ symmetry in the final correlation.

Next we turn to the other, apparently more curious,
correlations: the backward identity correlations
and the forward time-reversed one.
As an example of a backward $(l \le 0)$ identity correlation
consider $l=-1$:
\[ f^{-1}(q)= \left\{ 
\begin{array}{ll}
q/2 & \mbox{for } p < 1/2 \\
(q+1)/2 & \mbox{for } p > 1/2 
\end{array} \right. \]
Therefore
\begin{eqnarray}
\tilde C_{I}(-1)&=&\int_{0}^{1} dq \, \int_{0}^{1/2} dp \, F(q,p) \cos(\pi q)
\nonumber \\
&& - \int_{0}^{1} dq \, \int_{1/2}^{1} dp \, F(q,p) \cos(\pi q)
\end{eqnarray}
In fact, since $\cos(\pi q)$ vanishes at 1/2, there is no discontinuity
in the full integral, but it is more difficult to evaluate (and to
approximate).  If one were to take the upper limits of the $p$ integrals
to be infinity, there would be errors at $p=1/2$.  However, this is not
terribly damaging, and tolerating a small discontinuity at this point
due to this approximation leads to:
\beq
\tilde C_{I}(-1)\, =\, \pm \frac{1}{\sqrt{2} N} \exp(-\pi/(8N)) 
\cos(\pi q_{\alpha})
\eeq
the sign depending on if $p_{\alpha} <1/2$ or if $p_{\alpha}>1/2$
respectively.  The time-reversed, forward correlation, $\tilde C_{TR}(1)$,
is the same as this except for interchanging the roles of $q_{\alpha}$ and
$p_{\alpha}$.

The generalization of this to higher times is (take $l> 0$ below):
\beq
\tilde C_{I}(-l)\, =\, \sum_{\nu=0}^{2^l-1}\, \int_{0}^{1} dq \, \int_{\nu/2^l}
^{(\nu+1)/2^l}dp \, F(q,p) \cos(2 \pi ( q+\overline{\nu})/2^l)
\eeq
where $\nu$ represents a partition of the bakers map at time $l$,
and $\overline{\nu}$ results from the bit-reversal of the binary expansion
of $\nu$. The momentum gets exponentially partitioned with time, and
it precludes going beyond the log-time here as well
(like the forward correlation), although there is apparently no
super-exponential decrease here. Indeed if we evaluate the above after
neglecting finite limits in each of the $p$ integrals above, so that we
would have $2^l$ discontinuities at time $l$, we get:
\beq
\label{21}
\tilde C_{I}(-l)\, =\,
\frac{1}{\sqrt{2} N} \exp(-\pi /(2^{(2l+1)} N)) \cos(2 \pi (q_{\alpha}
+\overline{\nu})/2^l)
\eeq
depending on if $p_{\alpha}$ lies in the interval $(\nu/2^l, (\nu+1)/2^l)$.
So that for $l$ large and $N$ fixed, the exponential goes to unity;
effectively, for large $N$ and any $l$, the exponential can be replaced by
unity.  Even the $q_\alpha$-dependent part of the argument in the function
(cos) itself is tending to vanish, so that the integral seems to give the area
of the Gaussian ($h$).  The approximation of putting all $p$ limits to
infinity makes sense only if the Gaussian state is well within a zone of
the  partition and this is necessarily violated at half the log-time.  So
the approximate expression of Eq.~(\ref{21}) breaks down beyond
$\tau/2$.  This lack of a super-exponential cutoff as seen with the
previous correlations considered is due to two special conditions.
First, the argument of the cosine has no $p$-dependence.  Second,
all the stable manifolds are perfectly parallel to the $p$ axis.  We
would recover super-exponential decay in all the correlations if the
operator, $\hat A$, being considered was a constant function along
neither the stable nor the unstable manifold.  In this sense, we have
chosen a maximally difficult operator with which to test the semiclassical
theory, though it simplifies the quantum calculations.

As before, $\tilde C_{TR}(l)(q,p)=\tilde C_{I}(-l)(p,q)$.  Parity symmetry
is benign and leads to an overall multiplication by a factor of 2.  Thus,
the final semiclassical expression for the full correlation for the
quantum bakers map is:
\begin{eqnarray}
\label{bakfinal}
C_{A}(\alpha)& =& \frac{2}{N} \left[ \sum_{l=0}^{T_{1}}
\exp( -2^{2l} \pi /(2N)) \cos(2 \pi 2^l q_{\alpha}) \, \right. \nonumber \\
&& + \sum_{l=1}^{T_{2}}
\exp(-\pi /(2^{2l+1} N))
\sum_{\nu=0}^{2^l-1} \left( \cos(2 \pi (q_{\alpha}
+\overline{\nu})/2^l) \right. \nonumber \\
&& \left. \left. \times \Theta(p_{\alpha}-\nu/2^l)
\Theta((\nu+1)/2^l -p_{\alpha})\right) \right] \, \ \nonumber \\
&& +\, (q_{\alpha}
\leftrightarrow p_{\alpha}).
\end{eqnarray}
where $T_{1}$ can be infinite but it is sufficient to stop just beyond
half the log-time.  As just discussed, $T_{2}$, is more problematic here,
and we do not have an expression to use beyond $\tau/2$.  $\Theta$ is the
Heavyside step function that is zero if the argument is negative and unity
otherwise.  The correlation is of the order $1/N$ or $\hbar$.  If one were
to divide by the number of states in Eq. (\ref{correl}) so that it is a
true average, this quantity would decrease as $1/N^2$ or $\hbar^2$.

\begin{figure}
  \epsfig{file=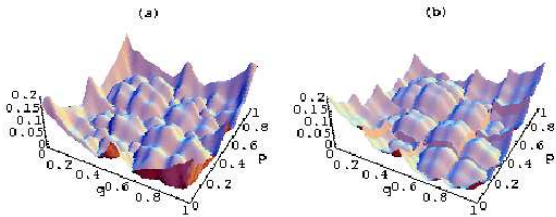, width=8.2cm, angle=180}
  \caption{(a) The absolute value of the
  quantum correlation with the $\cos(2 \pi q)$
  operator for $N=100$ whereas (b) corresponds to a
  semiclassical evaluation of the same.}
\end{figure}

For the case of $N=100$, we compare in Fig.~(1) the full quantum
correlation given by Eq.~(\ref{correl}) with the final semiclassical
evaluation given by Eq.~(\ref{bakfinal}).  The absolute value of the
correlation function is contoured and superposed on a grey scale.
Figure~(1a) shows the quantum calculation for the full phase space.  In
other words, the intensity (value) of each point, $(q,p)$, on the plot
represents the $C_A(\alpha)$-calculation for a wave packet centered at
$(q_\alpha=q,p_\alpha=p)$.  The first sum in Eq.~(\ref{bakfinal}) (over
$T_1$ terms) is a smooth function, and it also displays an additional
symmetry about $1/2$ in both canonical variables separately.  This extra
symmetry is broken by the second sum (over $T_2$ terms).  Figure~1(b)
compares the semiclassical formula to the exact quantum calculation.  We
have taken eight ``forward'' correlations (excluding zero), i.e. $T_1=8$,
while we have only taken two ``backward'' correlations, i.e. $T_2=2$.
This is because it appears that the approximations that go into the latter
expressions lead to non-uniformly converging quantities and it works
better to stop at a earlier point in the series.  The (artificial)
discontinuities at $1/2$ and $1/4$ are seen prominently in the
semiclassical results.  Otherwise, it turns out that the semiclassical
approximation captures many fine-scale features of the correlations, some
of which will be discussed below.  Figures~(2a,b) are for specific one
dimensional sections of the same quantities.  The agreement is very good.

\begin{figure}
  \epsfig{file=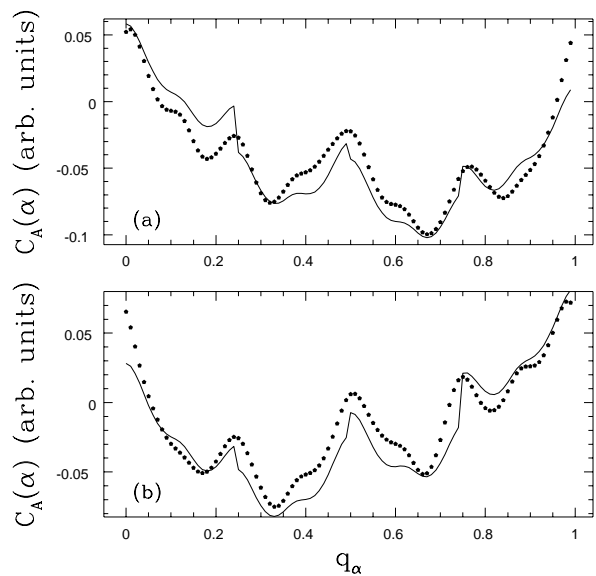, width=8.2cm}
  \caption{Sections of the correlation for $N=100$.
  (a) $p_\alpha=.33$ section, (b) $p_\alpha=.72$ section. The points
  are the quantum calculation while the solid lines are semiclassical
  evaluations.}
\end{figure}

\subsection {Classical features in the correlation}

A strong (positive) correlation is indicated at the classical fixed
points $(0,0)$ and $(1,1)$, with the rest of the significant correlations
being negative.   They are dominated by several classical structures as
illustrated in Fig.~(3).  Here the $N$ value used is 200, and superposed
on the significant contour features are the following classical orbits:

\begin{itemize}

\item[i)] the period-2 orbit at $(1/3,2/3)$, $(2/3,1/3)$ is by far the
most prominent structure. This is shown in Fig.~(3a) by two circular dots.
Also, we can look at these structures closely through 1-d slices.
In Fig.~(2a), the correlation is seen to be large and negative at
$(q_{\alpha}=.66)$.  The period-2 structure is dominating the
landscape;

\item[ii)] next in importance is the primary homoclinic orbit to the
period 2 orbit in i), $(1/3,1/3)$, which goes to $(2/3,1/6)$,
quickly gets into the region of the period two orbit and is difficult to
resolve.  The parity and time-reversal symmetric image points are also
indicated.  It turns out that there is an infinite set of periodic orbits
which approximate this orbit more and more closely.  Its effects may be
present simultaneously, and indistinguishable from the homoclinic
orbit itself~\cite{Sarac90}.  The relevant family (set) is denoted by
$(001)_{01}$ which is based on a complete binary coding of the
orbits~\cite{BalVor89}.  For example, the first few periodic orbits of
the family are associated with the binary codes
$(00101)$, $(0010101)$, and $(001010101)$.  They are also shown in
Fig.~(3b), including the symmetric image points. In the 1-d slice of
Fig.~(2a) we see this orbit as well;

\begin{figure}
  \epsfig{file=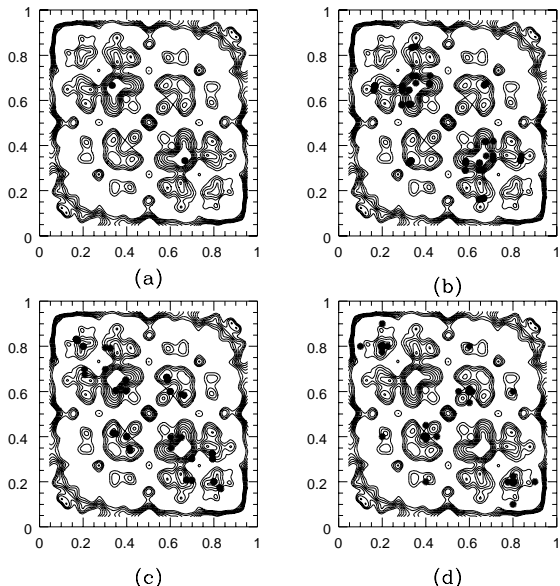, width=8.2cm}
  \caption{Comparison of classical structures in the correlation
  at $N=100$.  Details in the text.}
\end{figure}

\item[iii)] there is an infinite number of orbits homoclinic
to the period-2 orbit.  They become increasingly more complicated.  The
next associated periodic orbit family, $(0011)_{01}$, is shown in
Fig.~(3c), including the symmetric image points. This family was
noted by Saraceno to scar eigenfunctions~\cite{Sarac90}.  Also shown in
this figure is the period-4 along the diagonal lines: $(3/5,3/5)
\rightarrow (1/5,4/5) \rightarrow (2/5,2/5) \rightarrow (4/5,1/5)$.
Figure~(4) shows sections at $p_{\alpha}=3/5,4/5,2/5$ to highlight this
orbit.  In Figs.~(4a) $p_{\alpha}=3/5$ and has a local minimum at
$q_{\alpha}=3/5$; (b) $p_{\alpha}=4/5$ and has a local minimum at
$q_{\alpha}=1/5$; and (c) $p_{\alpha}=2/5$ and has a local minimum at
$q_{\alpha}=2/5$.  These are marked, to indicate location along
$q_{\alpha}$ by filled circles.  The other minima are due to competing
nearby structures of the period-2 orbit and its principal homoclinic
excursion;

\begin{figure}
  \epsfig{file=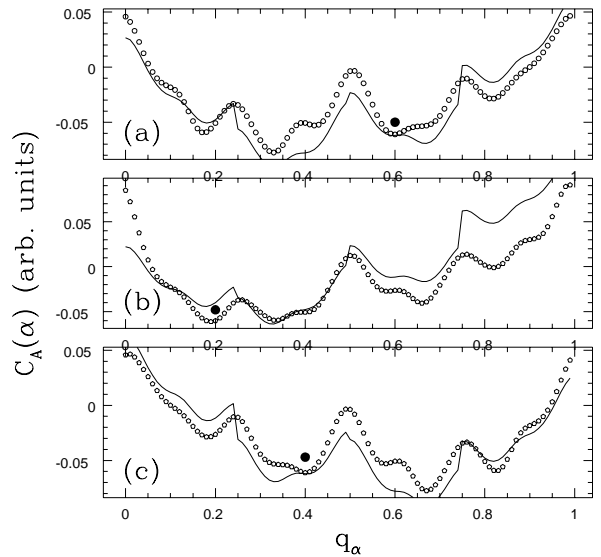, width=8.2cm}
  \caption{Sections of the correlation ($N=100)$ to highlight the
  period-4 orbit. (a) $p_{\alpha}=3/5$ and has a local minimum at
  $q_{\alpha}=3/5$ (b) $p_{\alpha}=4/5$ and has a local minimum at
  $q_{\alpha}=1/5$ (c) $p_{\alpha}=2/5$ and has a local minimum at
  $q_{\alpha}=2/5$.}
\end{figure}
	
\item[iv)] the orbit homoclinic to the period-4 orbit included in
Fig.~(3c) with the initial condition $(1/5,2/5)$ (and its symmetric
partners) is shown in Fig.~(3d); and

\item[v)] points, such as $(0,1/4)$, which are homoclinic to the fixed
points $(0,0)$, $(1,1)$ also show prominently.

\end{itemize}

\begin{figure}
  \epsfig{file=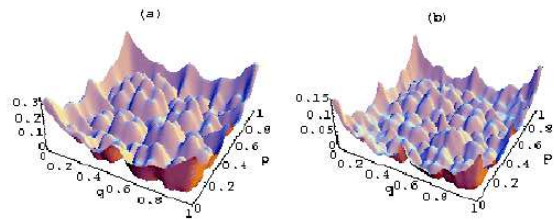, width=8.2cm, angle=180}
  \caption{The correlation at (a) $N=128$ and (b) $N=200$.  Note the
  sharp features in (a), where the peak height is about twice as large as
  that in (b).}
\end{figure}

That these structures are in a sense invariant, i.e. not specific to
$N=100$ is shown in Fig.~(5a,b) where the correlation (absolute value) is
shown for $N=128$ and $200$ respectively. The phase-space resolution of
the correlation  is increasing with $N$, while the overall magnitude is
decreasing as $1/N$. The peculiar properties of the quantum bakers map
for N equaling a power of two~\cite{BalVor89,Sarac90,OT91} is
tested by $ N=128$.  Here the correlation is ``cleaner'' and the
stable and unstable manifolds at 1/4, 1/2, and 3/4 of the fixed points
are clearly visible. The peaks are well enunciated as well.  Both
Figs.~(5a,b) have contours up to 2/3 peak height, so a  direct comparison
is meaningful. Higher $N$ values show more clearly the secondary
homoclinic orbit to the period-2 orbit.

We may compare these structures with the inverse participation ratio
defined as:
\beq
P(\alpha)\, =\,  \sum_{i=0}^{N-1} \left |\br \alpha|\psi_{i} \kt \right|^4.
\eeq
It is illustrated in Fig.~(6).  It shows marked enhancements at the
period-2 and period-4 (along the symmetry lines) orbits, and closer
examination reveals all orbits  up to period-4 are present and one orbit
of period-6 along the symmetry lines (the diagonals); see
Ref.~\cite{OT91} for a more detailed discussion.

\begin{figure}
  \epsfig{file=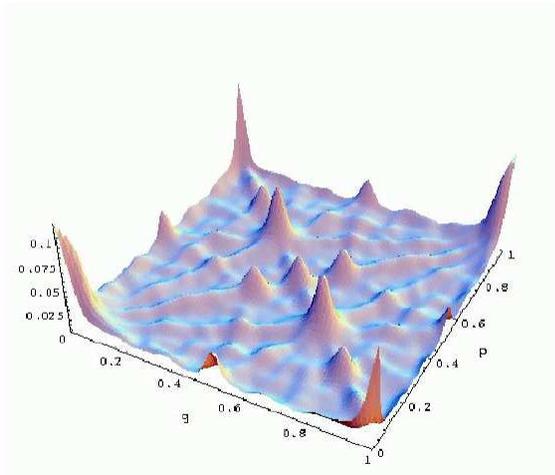, width=8.2cm, angle=180}
  \caption{The participation ratio for the bakers map.  Classical
  structures are present in this quantity as well.}
\end{figure}

\subsection{General operators and selective enhancements}

The results so far have dealt with the special case $A(q)=\cos(2 \pi q)$.
It seems natural to suspect that the structures highlighted in the
correlation are dependent on the choice of the operator.  This turns
out to be true, and we show here how this works in the bakers map.  We
reemphasize though that were the eigenstates behaving ergodically, the
correlations would have been consistent with zero to within statistical
uncertainties independent of the choice of the operator.  In this sense,
a complete view of the extent to which the eigenstates manifest phase
space localization properties comes only from considering both the full
phase plane of wave packets and enough operators to span roughly
the space of possible perturbations of the energy surface.  The
flexibility of operator choice does provide a means to enhance
selectively particular features of interest supposing one had a specific
localization question in mind.  As an illustration, note that
localization about the period-3 orbit barely appeared in the contour plot
of Fig.~(3), and yet, we show below that it can be made to show up
prominently with other operators.

\begin{figure}
  \epsfig{file=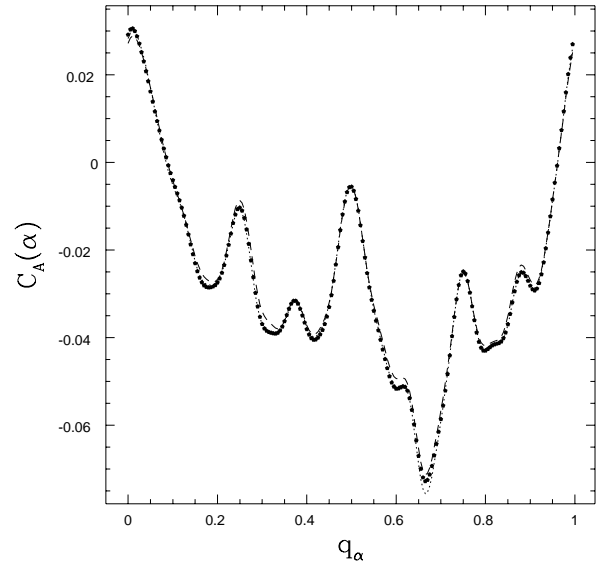, width=8.2cm}
  \caption{The correlation for the series of operators
  $A=cos( 2\pi 2^m q)$. Shown are $m=0$ (large dot),  $m=1$ (small dot),
  $m=2$ (dashed line).}
\end{figure}

Since the case $A(q)=\sin(2 \pi n q)$ has vanishing correlations for any integer
$n$ due to symmetry, the other cases of interest are the higher
harmonics of the cosine. Therefore consider:
\beq
	A(q)=\cos(2 \pi n q).
\eeq
If $n=2^m$ for some positive integer $m$, a rather remarkable scaling
property of the quantum bakers map is revealed that is actually implicit
in the way the bakers map was originally quantized in \cite{BalVor89}.
Semiclassically, the correlations are {\em identical} to the case $m=0$.
For example, consider $A(q)=\cos(4 \pi q)$.  Then the one-step back classical
correlation becomes identical to the zero-th order correlation
corresponding to $A(q)=\cos(2 \pi q)$. The correlations all shift by $m$ in
the sense that $C(l) \longrightarrow C(l+m)$.   Thus, there is a kind of
scale invariance in the correlation like classical fractals, although
this is not self-similarity in the same curve.  Quantum calculations
reflect this invariance to a remarkable degree as seen in Fig.~(7)
where the $N=200$ and $p_{\alpha}=1/3$ case is shown.

Other harmonics do weight differently the same localization effects
(classical structures).  In Fig.~(8), $N$ and $p_{\alpha}$ are taken
the same as in Fig.~(7).  The cases $n=1,3,5$ are all very different
from each other, but note that the case $n=6$ almost coincides with
$n=3$ for the same reason that powers of two harmonics are nearly same.
Thus only operators of odd harmonics give the possibility of providing
new or unique information about the nonergodicity in the eigenstates.  The
period-2 orbit localization is accentuated at $n=3$, since for $n=3m$ where
$m$ is a positive integer, $\cos(6 \pi m q)$ has a maximum of $+1$ at
$q = 2/3$, whereas for all other integers $n$, $\cos(2 \pi n q) = -1/2$ 
at the same point.  In short the perturbation (or measurement) is
more significant at the location of the period-2 orbit for $n=3$.
On the other hand, the case $n=5$ is similar to the fundamental harmonic 
case at $q = 2/3$ where the perturbation is also equal to $-1/2$.

\begin{figure}
  \epsfig{file=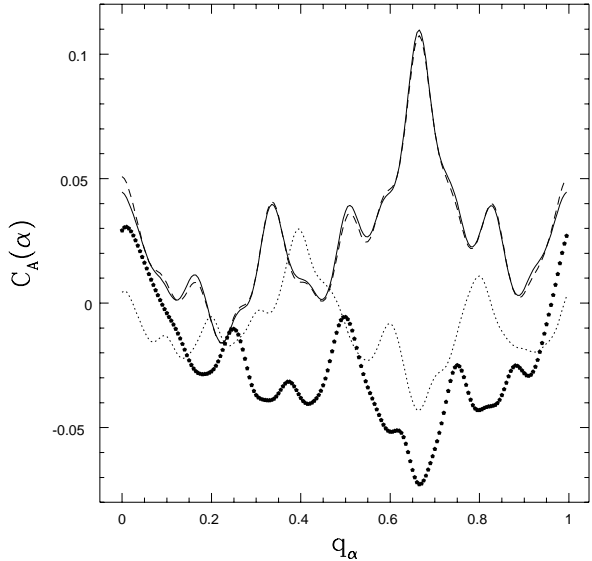, width=8.2cm}
  \caption{Correlation for different harmonics $A=cos(2 \pi n q)$.
  Shown are $n=1$ (solid large dots), $n=3$ (solid line), $n=5$ (dotted
  line),  and $n=6$ (dashed line).}
\end{figure}

The case $n=7$ is interesting as $\cos(14 \pi q)$ has a maximum at 
$q=1/7$ which coincides with a period-3 orbit at $(1/7,4/7)$.  
In Fig.~(9), we see the correlation ($N=100)$
corresponding to this operator and the dominant structure is this
period-3 orbit and its symmetric partner.  Also visible are the stable
and unstable manifolds of these orbits.  In fact, it is the multiples
of the $2^m - 1$ harmonics which selectively highlights the period
$m$ orbits.

Summarizing then, the correlations reflect that the bakers map
eigenstates are not ergodic, and manifest strongly phase space
localization properties.  There do not exist transport barriers such as
cantori or diffusive dynamics in the bakers map, so whatever localization
that exists should be due to scarring by the short periodic orbits.  This
is confirmed in the examples shown with connections to their homoclinic
orbits illustrated as well.  The perturbation or observable determines
the regions of phase space that will light up in the correlation
measure.  A semiclassical theory predicts reasonably well many of
these structures.  The correlation is semiclassically written as a sum of
classical correlations that are super-exponentially cut off after about
half the log-time scale.

\vspace{-1.5cm}
\begin{figure}
  \epsfig{file=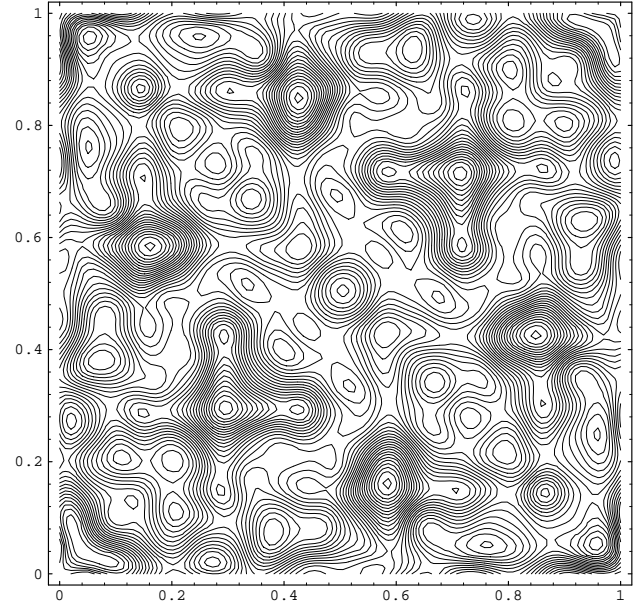, width=8.2cm}
  \vspace{-1.5cm}
  \caption{The correlation for the operator $A=cos(14 \pi q)$ and
  case $N=100$.  The highlighted areas are in the region of classical
  period-3 orbits.}
\end{figure}

\section{ The standard map}

\subsection{The map and the mixed phase  space regime}

The standard map (a review is found in~\cite{Izrailev}) has many
complications that can arise in more generic models and we turn to their
study. It is also an area-preserving, two-dimensional map
of the cylinder onto itself that may be wrapped on a torus.  We will
consider identical settings of the phase space  and Hilbert space as for
the bakers map discussed above. The standard map has a parameter that
controls the degree of chaos and thus we can study  the effect of regular
regions in phase space, {\em i.e.} the generic  case of mixed dynamics.

The classical standard map is given by
the recursion
\begin{eqnarray}
\label{classmap}
q_{i+1}&=& (q_{i}+p_{i+1}) \, \mbox{mod}\,  (1)\\ \nonumber
p_{i+1}&=&(p_{i}-(k/2 \pi) \sin(2 \pi q_{i})) \, \mbox{mod}\,  (1),\nonumber
\end{eqnarray}
where $i$ is the discrete time. The parameter $k$ is of principal
interest and it controls the degree of chaos in the map.  Classically
speaking, an almost complete transition to ergodicity and mixing is
attained above values of $k \approx 5$, while the last rotational KAM
torus breaks around $k
\approx .971$.

The quantum map in the discrete position basis is given by \cite{LakPRM}
\begin{eqnarray}
\label{quantummap}
       \br n |U| n^{\prime} \kt \, &=&\, \frac{1}{\sqrt{i N}}
\exp\left(i \pi (n-n^{\prime})^2/N\right) \nonumber \\
&& \times \exp \left(i \frac{k N}{2 \pi} \cos(2 \pi (n+a)/N) \right).
\end{eqnarray}
The  parameter to be varied will be the ``kicking strength'' $k$, 
while the phase
$a=1/2$ for maximal quantum symmetries, and $n, n^{\prime}\, =\,
0, \ldots, N-1$.

We use the unitary operator and evaluate the correlation as in
Eq. (\ref{correl}) with $A(q)=\cos(2 \pi q)$ here as well. This
corresponds exactly to the level velocity induced by a change in
the parameter $k$. In Fig.~(10) is shown the absolute value of the
correlation for various values of parameter $k$. Case (a) corresponds
to $k=.1\times(2 \pi)$ and is dominated by the KAM curves as the
perturbation has not yet led to significant chaos.  Highlighted is the
fixed point resonance region at the origin that is initially
stable.  An unstable point is located at the point $(1/2,0\ {\rm or}\
1)$.  The separatrix or the stable and the unstable manifolds of this
point are aligned along the local ridges seen in the correlation.  Also
the period-2 resonance region is visible.  Higher resolution not shown
here, corresponding to higher values of $N$ reveal weakly the period-3
resonance as well.  Case (b) corresponds to $k=.3 \times(2 \pi)$ while
(c) and (d) to $k=.9 \times (2 \pi)$ and $k=2.3 \times (2 \pi)$
respectively.  We note the gradual destruction of the KAM tori and the
emergence of structures that are dominated by hyperbolic orbits. A more
detailed classical-quantum correspondence is however not attempted here.

\begin{figure}
  \epsfig{file=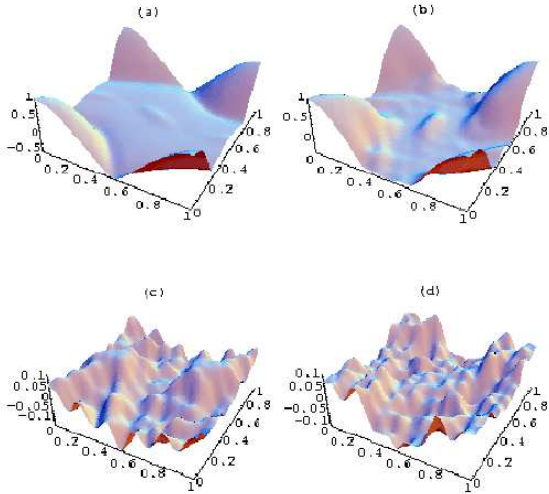, width=8.2cm, angle=270}
  \caption{The quantum correlation for the standard map ($N=100$),
  (a) $k/(2 \pi)=.1$, (b) $k/(2 \pi)=.3$, (c) $k/(2 \pi)=.9$,
  (d) $k/(2 \pi)=2.3$.}
\end{figure}

These contour plots do not reveal the difference in the magnitude
between the correlations in the stable and unstable regions.  In
Fig.~(11), we have plotted the correlation at the origin $(0,0)$, which
is also a fixed point, as a function of the parameter.  The value $k/(2
\pi) \sim 1$ corresponds to a transition to complete classical chaos
and is reflected in this plot as erratic and small oscillations. The
large correlation in the mixed phase space regime arises from the
non-ergodic nature of the classical dynamics.  The classical fixed
point loses stability at $k^{\ast}/(2 \pi)=4/(2 \pi)\approx .63$ and
this is roughly the region at which the correlation starts to dip away
from unity toward lower values.

\begin{figure}
  \epsfig{file=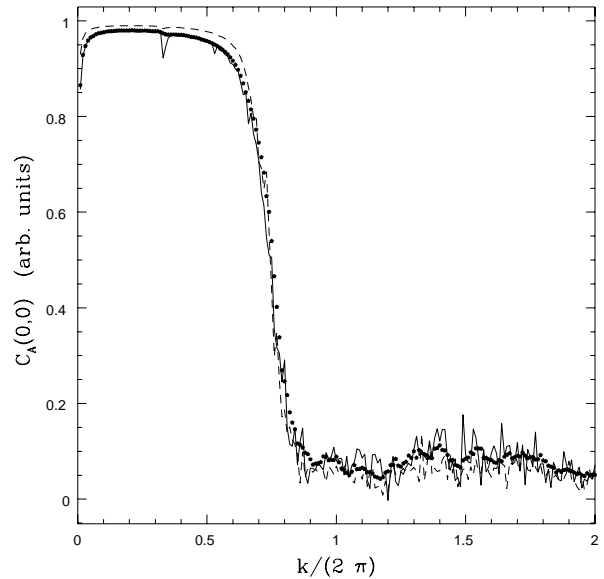, width=8.2cm}
  \caption{The quantum correlation for the standard map
  at $\alpha=(0,0)$ as a function of the parameter $k$ ($N=100$ is the
  solid line and $N=200$ is the dashed one).  The dotted line is a classical
  estimate based on Eq.~(\ref{classreg}) and averaged over twenty  time
  steps with $N=100$.}
\end{figure}

The gross features and principal $\hbar$ behavior in this regime is easy
to derive in terms of purely classical correlations as follows:
\begin{eqnarray}
C_A(\alpha)\, &=&\, \left<\mbox{Tr}\left(|\alpha\kt \br \alpha|\hat{A}(n)\right)
\right>_n\, \nonumber \\
&=&\, \left< \int dq\, dp \, [|\alpha\kt \br \alpha|]_W \, [\hat{A}(n)]_W \right>_n
\end{eqnarray}
where $[.]_W$ is the Weyl-Wigner transform of the operator in the
brackets and $\hat{A}(n)$ is the operator after a time $n$.  Without worrying
about the toral nature of the phase space and the Weyl-Wigner transforms,
we treat the problem as in a plane. This is justified by the use
of localized, Gaussian wave packets.  Otherwise, we could imagine that
the Wigner transform of the projector would follow from an infinite
series of Gaussian states that is equivalent to discretizing the
Gaussian.  We use a normalized, ``circular'' Gaussian of
width $\sqrt{\hbar}$.  The approximation comes in when we replace
$[\hat{A}(n)]_W$ by $A(f^n(q,p))$ where the latter is the classical function
evaluated at the classically iterated point
$q_n=f^n(q,p)$.  We expect this approximation to be valid in the
case of regular dynamics over a much longer time scale than found with
chaotic dynamics.  To a good approximation,
\begin{eqnarray}
\label{classreg}
C_A(\alpha)\, &\sim& \,\left< \int dq\, dp \, \left(\frac{1}{\pi \hbar}\right)
\exp\left[-\frac{1}{\hbar}\left((q-q_\alpha)^2 \right. \right. \right. \nonumber \\
&& \left. \left. \left. +(p-p_\alpha)^2 
\right)\right] \, A\left(f^n(q,p)\right)\right>_n
\end{eqnarray}
As intuitively expected, there is no principal $\hbar$ dependence
in the correlation since there is a non-zero classical limit.  At
$\hbar=0$, we could replace the Gaussian forms by $\delta-$functions and
would get simply $C_A(\alpha)\, =\, \left<
A(f^n(q_\alpha,p_\alpha))\right>_n$.

This however vanishes as the classical system becomes more ergodic and is
no more capable of predicting the correlation.  Higher-order
corrections are needed. It is in this regime that we studied the bakers
map and found that the correlation has a principal part that scales
(almost) as $\hbar$ and classical correlations based on periodic
orbits predict the localization features that arise out of quantum
interference.

We return to Fig.~(11) to remark on some of these properties.  Notice that
the simple estimate of Eq.~(\ref{classreg}) performs very well, even as
the phase space is becoming increasingly chaotic.  It is quite unexpected
that the oscillations after the onset of full mixing (around $k/(2 \pi)
=1$) should follow this estimate.  However, after the transition to chaos
the classical estimate will depend on the times over which the averaging
is done and as this increases the estimate would vanish.

\subsection{Chaotic regime}

We attempt in some measure a semiclassical theory for the correlation
in the chaotic regime along the lines adopted for the quantum bakers map.
Of the two ingredients in Eq. (\ref{obscure}) one of them remains
the same, namely Eq. (\ref{semtrace}). However the diagonal elements
of the propagator in Eq. (\ref{semprop}) have to be generalized.

In~\cite{SteveHell93} a semiclassical expression for the matrix elements
of the propagator as a homoclinic orbit sum is given.  Although this was
derived with the example of the billiard in mind, it can be interpreted as
a generalization of Eq.~(\ref{semprop}) for area-preserving,
two-dimensional maps.  We, however, interpret the sum not as a homoclinic
orbit sum, but as a periodic orbit sum.  To each homoclinic orbit there
is a neighboring periodic orbit that we will use instead. This will form
the points around which the expansions are carried out and the result is
identical to that in \cite{SteveHell93}. Thus we write
\beq
\label{sempropg}
\br \alpha|U^n|\alpha \kt \, \sim \, \sum_{\gamma}
\exp\left(i S_{\gamma}/\hbar -i \pi \nu/2\right )\sum_{j} B_j
\eeq
where
\begin{eqnarray}
\label{B}
B_j\, &=&\, \sqrt{\frac{2}{A_0}}\exp\left\{\frac{-1}{2 \hbar A_0}
\left[(\mbox{tr}-2)(\delta q^2+ \delta p^2)\, \right. \right. \\
&& \left. \left. +\, 2 i (m_{21}\, \delta p^2-
m_{12}\, \delta q^2+\delta q\delta p\, (m_{22}-m_{11})) \right] \right\}
\nonumber.
\end{eqnarray}
$B_j$ generalizes the Gaussian form (including the prefactor) in
Eq.~(\ref{semprop}).  Again $j$ labels points along the periodic orbit
and $\delta q=q_j-q_{\alpha}$, $\delta p=p_j-p_{\alpha}$ are as before
deviations from the centroid of the wave packet.  The two dimensional
matrix elements, $m_{ij}$, are the elements of the stability matrix
at the periodic point $j$ along the periodic orbit $\gamma$. 
The  deviations $\delta q$ and $\delta p$ after $n$ iterations of the map are
given by:
\beq
\label{monodrom}
\left( \begin{array}{c}
	\delta p_n\\ \delta q_n \end{array}\right)
\,=\, \left( \begin{array}{ll}
	m_{11} & m_{12} \\ m_{21}&m_{22} \end{array}\right)
\left( \begin{array}{c}
	\delta p\\ \delta q \end{array}\right)
\eeq
The invariant
is the trace of this matrix that is denoted tr. While
$A_0\, =\, m_{11}+m_{22}+i (m_{21}-m_{12})$,  $\nu$ is a phase that
will not play a crucial role below.
In the case of the bakers
map $m_{12}=m_{21}=0$ and $m_{11}=2^{-n},\,  m_{22}=2^{n}$ uniformly at
all points in phase space, as well as $\nu=0$.
On substitution of this in Eq. (\ref{sempropg})
we get Eq. (\ref{semprop}).

The dependence on individual matrix elements of the stability matrix
complicates the use of this formula in general. However we note that the
Gaussian is effectively cutting off periodic points that are not close to
$\alpha$ and therefore we may take the $m_{ij}$ elements to be the
stability matrix at this point.  In the chaotic regime each of the matrix
elements   grow exponentially with time $n$. Thus we have  that
$\exp(-\lambda n) \, m_{ij} \rightarrow$ constant, where $\lambda$
is the Lyapunov exponent. We call this saturated constant $m_{ij}$
as well. Below we will assume that the exponential growth has been
factored out of these elements. Also we use $ \exp(-\lambda n) \,A_0
\rightarrow a_0$. The terms inside the exponential function in
Eq. (\ref{B}) saturate in time $n$ while the prefactor goes as
$\exp(-\lambda n/2)$. It follows then  that
\beq
B_{j} \rightarrow \sqrt{\frac{2}{a_0}} \exp(-\lambda n/2) \, F(q_j,p_j)
\eeq
where $F(q_j,p_j)$ is
\begin{eqnarray}
\label{F}
F(q_j,p_j)\, &=&\,
\exp\left\{\frac{-1}{2 \hbar a_0}
\left[(\delta q^2+ \delta p^2)\, \right. \right. \\
&+\,& \left. \left.  2 i (m_{21}\, \delta p^2-
m_{12}\, \delta q^2+\delta q\delta p\, (m_{22}-m_{11})) \right] \right\}.
\nonumber
\end{eqnarray}
Here the $m_{ij}$ elements already have the exponential behavior factored
out.  For example in the case of the bakers map $m_{22}=1$ while all the
other elements are zero and this gives consistently the approximated
Gaussian form in Eq. (\ref{bakerF}).
Further steps are identical to the case of the bakers map and leads to the
generalization of Eq. (\ref{fincor}):
\beq
\label{fincorg}
C_A(\alpha)\, =\, \sqrt{\frac{2}{a_0}}
\sum_{T} \sum_{l= -M }^{M} \tilde C_T(l),
\eeq
where the classical correlations are calculated as in Eq. (\ref{classcor})
with the function $F$ being that in Eq. (\ref{F}).
We may then expect all the principal conclusions from the study of the
bakers map to be carried over, principally the decrease in the
correlation as $\hbar$, the correlations being cut off after half the
log-time scale, and the effects of classical orbits.

More detailed analysis in the lines of the special case discussed in
case of the bakers map will run into the following difficulties.  First,
the $m_{ij}$ elements will depend on $\alpha$ in general.  Exceptions
are uniformly hyperbolic systems such as the cat or sawtooth maps
(and, of course, the bakers map).  A second difficulty is that the
correlations have to be evaluated to half the log-time while classically
iterating the map (analytically) over such times is often not possible.
The classical correlations that arise in the study of rms values of
level velocities~\cite{ANS99} involved correlations that exponentially
decreased in time while here we are likely to get generalizations of
forms such as in Eq.~(\ref{forcor}) that will require us to go up to
log-times.   We have calculated the correlations for times 1, -1, and 2
but will not display them as they are by themselves not very useful.
A third problem with this form of the generalization is that it is not
explicitly real.

\begin{figure}
  \epsfig{file=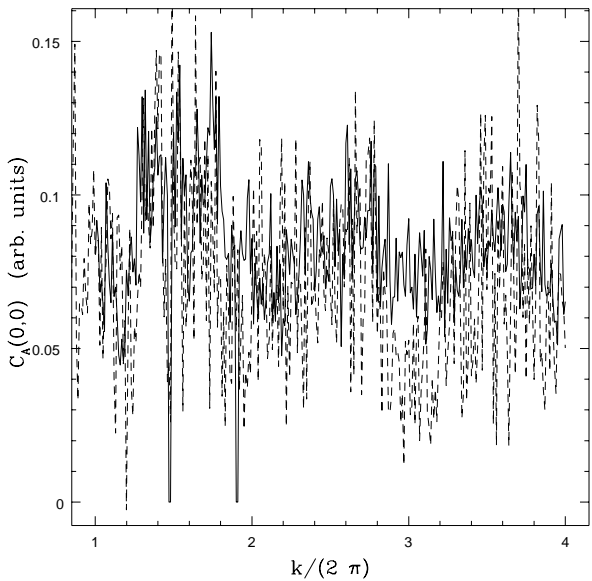, width=8.2cm}
  \caption{The quantum correlation for the standard map ($N=100)$
  at $\alpha=(0,0)$ as a function of the parameter $k$ (dashed line).  The
  solid line is the semiclassical estimate.}
\end{figure}

We have used Eq.~(\ref{fincorg}) and for the $m_{ij}$
used either those calculated at one point in phase space (such as
the origin) or in fact assumed those that are relevant for the bakers
map.  While fine structures are not reproduced, the general features
are captured equally well in both these approaches.  To illustrate
the quality of the approximation we again look at the correlation
at the point $(0,0)$ as a function of $k$ in Fig.~(12) (as in the
previous figure).  The solid line is the semiclassical prediction based
upon using the {\em same} $m_{ij}$ values at all values of $k$. It is seen
that even with these (over) simplifications the semiclassical expressions
capture much of the oscillations with the parameter and the magnitude.

\section{Summary and conclusions}

We have studied the details of phase space localization present
in the quantum time evolution of operators. This was related to a measure
of localization involving the correlation between the level velocities
and wavefunction intensities. While individual quantum states show
well known interesting scars of classical orbits, groups of states
weighted appropriately provide both a convenient and important
quantity to study semiclassically.
We were interested principally in those
features whose origins were quantum mechanical. The quantities studied
had both a vanishing classical limit as well as vanishing RMT averages.

We studied simple maps as a way to understand
the general features that will appear. We found that the operator dictated
to a large extent which parts of phase space will display prominent
localization features and further that these localization features are often
related to classical periodic orbits and their homoclinic structures.
The time average of the operator for wave packets was explicitly
related semiclassically to classical correlations. These  were shown
to be cut off super-exponentially after half the log-time in the quantum
bakers map. Thus the localization features in quantum systems associated
with  scars were reproduced using long (periodic) orbits but short time
correlations. The localization would disappear in the classical limit
as the magnitude of the quantum correlations or time averages are
proportional to (scaled) $\hbar$.

General systems were approached using the quantum standard map and
complications that would arise were discussed. Also the case of mixed
phase space was seen to be well reproduced by a simple classical
argument.  The generalization to Hamiltonian systems~\cite{Nick}
contains many of the features and structures are also
(not surprisingly) present in this case.

We gratefully acknowledge support from the National Science
Foundation under Grant No.~NSF-PHY-9800106 and the Office of Naval 
Research under Grant No.~N00014-98-1-0079.

\end{document}